\documentclass{ifacconf}

\usepackage{graphicx}      
\usepackage{epsfig} 
\usepackage{mathptmx} 
\usepackage{times} 
\usepackage{amsmath} 
\usepackage{amssymb}  
\usepackage[utf8]{inputenc}
\usepackage{subfigure}
\usepackage{float}
\usepackage{changepage}
\usepackage{algorithm}
\usepackage{multirow}
\usepackage[T1]{fontenc}
\usepackage{vmargin}
\usepackage{textcomp}

\usepackage{natbib} 

\begin{document}
\setcitestyle{numbers,square,comma,}
\begin{frontmatter}
\centering
\title{On the interplay between turbulent forces and neoclassical particle losses in Zonal Flow dynamics} 

\author{R. Gerrú$^{1,2}$,} 
\author{S. Mulas$^1$,} 
\author{U. Losada$^1$,}
\author{F. Castejón$^1$,}
\author{B. Liu$^{3,4}$,}
\author{T. Estrada$^1$,}
\author{B Ph van Milligen$^1$,}
\author{C. Hidalgo$^1$}

$^{1}$Laboratorio Nacional de Fusión, CIEMAT, 28040 Madrid, Spain\\
$^{2}$Departamento de Física Teórica, Universidad Autónoma de Madrid, 28049, Madrid, Spain\\
$^{3}$Hebei Key Laboratory of Compact Fusion, Langfang 065001, China\\ 
$^{4}$ENN Science and Technology Development Co., Ltd., Langfang 065001, China

\begin{abstract}            
This study presents the investigation of the connection between radial electric field, gradient of Reynolds stress and Long Range Correlation (LRC), as a proxy for Zonal Flows (ZF), in different plasma scenarios in the TJ-II stellarator. Monte Carlo simulations were made showing that radial electric fields in the range of those experimentally measured have an effect on the neoclassical orbit losses. The results indicate that, despite the order of magnitude of turbulent acceleration is comparable to the neoclassical damping of perpendicular flows, its dependence with radial electric field is not correlated with the evolution of LRC amplitude, indicating that turbulent acceleration alone cannot explain the dynamics of Zonal Flows. These results are in line with the expectation that the interplay between turbulent and neoclassical mechanisms is a crucial ingredient of the dynamics of edge Zonal Flows.
\end{abstract}
\begin{keyword}
Zonal Flows, Momentum transference, turbulence, neoclassical losses.\\
\\
(Some figures may appear only in colour in the online journal)
\end{keyword}

\end{frontmatter}

\section{Introduction}
The resistance of fluids to shearing motion is a well-known phenomena, where the tendency of sheared motion to be reduced with the passage of time leads to the concept of a positive coefficient of viscosity. However, for certain kind of flows (e.g. planet\textquotesingle s atmosphere and plasmas) evidence of negative viscosity effects has been reported; in this case the mean flow can gain energy from the turbulence with direct impact in the development of sheared flows with Zonal Flow (ZF) structure \cite{Diamond_review,Fujisawa_review}. Although it is known that ZF structures gain energy from turbulence, with direct impact in the regulation of plasma turbulence energy and particle losses, it is not clear which plasma conditions should be fulfilled to produce the development of these structures.\\
\\
Stellarator devices have pioneered the detection of Long-Range Correlations (LRC), consistent with the theory of Zonal Flows \cite{Fujisawa}. Later, experiments have shown that LRC in potential fluctuations are amplified either by radial electric fields externally imposed \cite{Pedrosa,Wilcox, Manz, Xu} or amplified by ambipolar radial electric fields (in L mode) \cite{LosadaL} or when approaching the L-H confinement edge transition \cite{Hidalgo,LosadaL-H}.\\
\\
Search for the causal relation between Zonal Flows, Reynolds stress (RS) and transport has been reported \cite{Birkenmeier, Alonso, Tynan}. In particular, experimental \cite{Birkenmeier,Schmid} and simulation \cite{Manz2} evidence of strong poloidal RS asymmetry points out that care should be taken to prevent a misleading interpretation of local RS measurements, as the ones presented in this study. Furthermore, the interplay between turbulent and neoclassical mechanisms has been identified during transitions in stellarators \cite{Velasco} and tokamaks \cite{Chang}.\\
\\
Those findings illustrate how unique diagnostic capabilities, to detect both Long Range Correlation (LRC) and to measure RS, are needed for deeper understanding of ZF dynamics.\\
\\
This paper addresses the mechanisms underlying the observed interplay between neoclassical radial electric fields and the amplification of low frequency zonal-flow-like structures \cite{Pedrosa}. With this goal, we have measured simultaneously the impact of radial electric fields ($E_r$) on LRC, gradients of Reynolds stress and simulated the influence of $E_r$ on neoclassical particle losses in the TJ-II stellarator. 

\section{Experimental Set-up and Simulation tools}
The equality of ion and electron fluxes (i.e. the ambipolar condition) determines the radial neoclassical electric field, which has two stable roots in stellarators: the ion root with typically negative $E_r$, usually achieved in high density plasmas with NBI heating, and the electron root with positive $E_r$, that is typically realized when electrons are subject to strong ECRH heating.\\
\\
Experiments were carried out in TJ-II stellarator, in both electron and ion root, each one in phases with pure ECRH heating (480 kW) and pure NBI heating (840 kW) respectively, and near to the transition from electron to ion root with ECRH heating. TJ-II has a toroidal magnetic field $B\approx1\>T$ and a minor radius of $a\approx0.2\> m$. The experiments were accomplished in the standard magnetic configuration, which has an edge rotational transform close to 1.6 such that the 8/5 rational surface is located at $\rho\approx0.8$.

\begin{figure}
    \centering
    \textbf{\hspace*{-0.5cm}}\includegraphics[width=8cm]{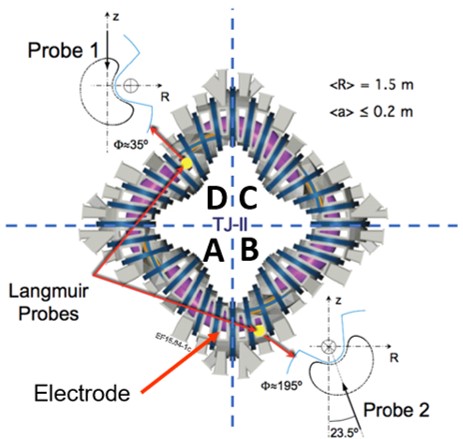}
    \caption{TJ-II scheme with the position of Langmuir probes and electrode \cite{LosadaL}.}
    \label{fig:TJ-II}
\end{figure}

The edge radial electric field was modified by two means: through a biased electrode (Fig. \ref{fig:TJ-II}) inserted into the plasma edge ($\rho\approx0.85$) and exploring plasma scenarios in the electron and ion roots with positive and negative radial electric fields respectively. The applied biasing was a low frequency triangular shape signal with an amplitude of about $\pm$400 V.\\
\\
A set of two Langmuir probes arrays located in sectors D and B (probe 1 and 2 respectively in Fig. \ref{fig:TJ-II}) were used to measure floating potential, Reynolds stress gradients and Long Range Correlation. Both probes are installed on a fast reciprocating drive on top and bottom of the device respectively. Probe D is positioned perpendicular to the magnetic field and consist of 5 rows of pins arranged in a stair-like shape, separated 5mm in radial direction and 3 mm in the parallel one, with 4 pins in each row separated by 3mm in perpendicular one, which allows 2-D measurements of plasma profiles and fluctuations. These pins are configured to measure floating potential ($\phi_f$) or ion saturation current ($I_{sat}$). Present experiments were done with the pins that measure $\phi_f$ of the three first rows, as shown in figure \ref{fig:RS}. Probe B has a rake form and consists of 12 pins in radial direction with a separation of 3mm and three poloidally separated pins at the front edge, again measuring $\phi_f$. The sampling rate of the floating potential signals is 2 MHz.

\subsection{Data analysis}
\subsubsection {Long Range Correlation (LRC) \newline}

Previous experiments have shown that LRC are dominated by frequencies below 20 kHz \cite{Silva}. Therefore, a low pass filter (1 - 20 kHz) was applied to floating potential signals from probe B and D. These filtered signals were used to compute the amplitude of LRC between them, without any time delay, as can be seen in equation \ref{eqn:correlation}. The computation of correlation was done with 2 ms time-windows.

\begin{equation}
    \gamma_{B-D}=\int_{-\Delta\tau/2}^{\Delta\tau/2}{\frac{\Phi_{f(B)}(t)\Phi_{f(D)}(t)}{\sqrt{(\Phi_{f(B)}(t))^2(\Phi_{f(D)}(t))^2}}dt}
    \label{eqn:correlation}
\end{equation}

\begin{figure}[]
    \centering
     \textbf{\hspace*{-0.5cm}}\includegraphics[width=7cm,trim={0 0 15cm 0}]{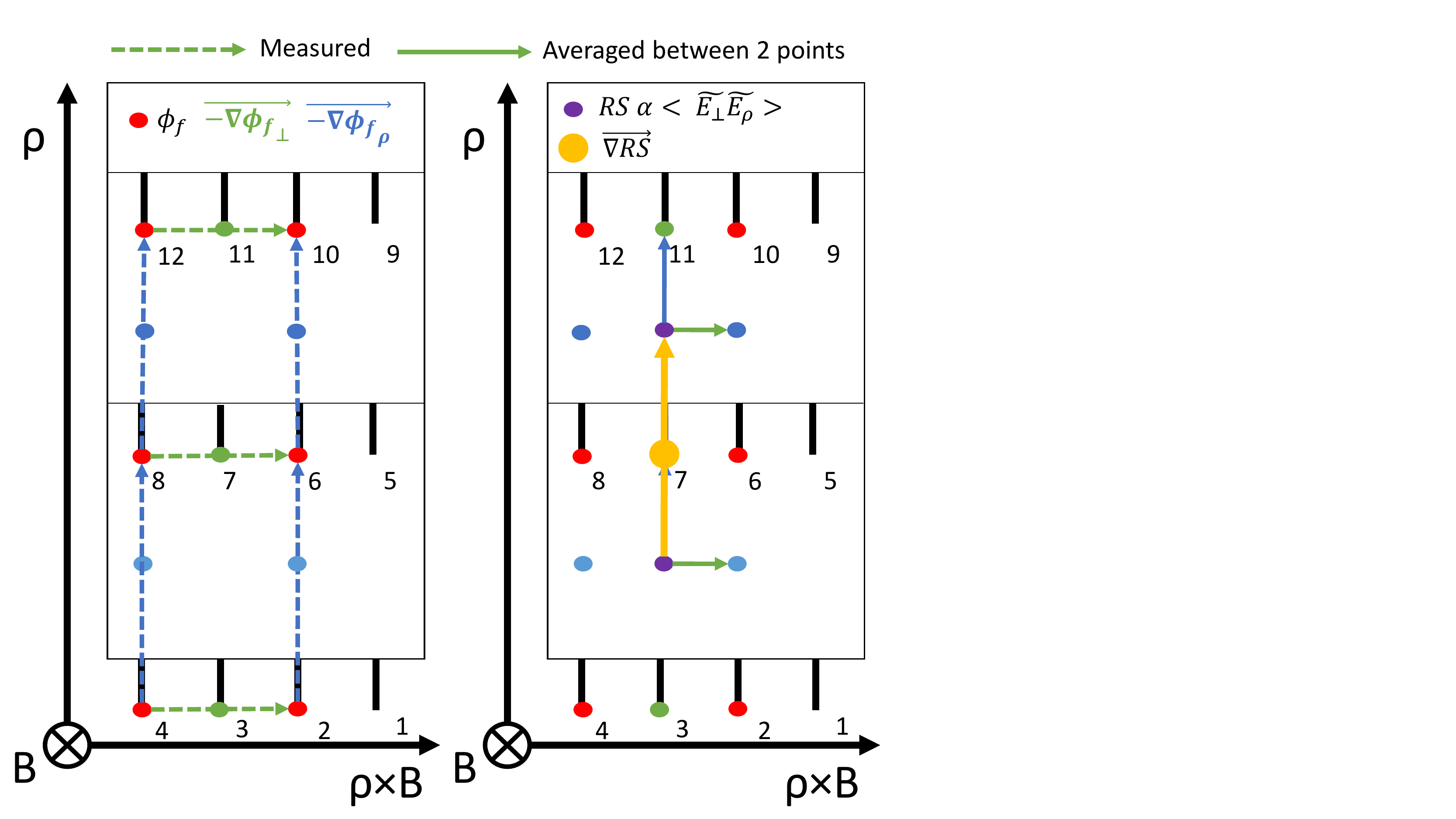}
    \caption{Scheme of probe D measurements of $\phi_f$ and the geometric configuration to compute $\nabla RS$.}
    \label{fig:RS}
\end{figure} 

\subsubsection{Gradient of Reynolds stress ($\nabla RS$)\newline}

Gradient of Reynolds stress computation was done using the high frequency fluctuations of the $\phi_f$ signals, from 20 to 250 kHz (studies exploring a different frequency ranges, such as 20 - 500 kHz and 10 - 500 kHz, have been done with no relevant difference in results). Using the pattern of probe D as is shown in figure \ref{fig:RS}, the fluctuating electric field in radial and perpendicular direction ($\tilde{E_r}$ and $\tilde{E_\theta}$) can be obtained, neglecting the influence of electron temperature fluctuations. These fluctuating electric fields induce a fluctuation in velocities (as $v_r$ and $v_\theta$), due to the E$\times$B drift, which were computed in order to quantify RS (\ref{eqn:RS}) and its gradient in radial direction as displayed in figure \ref{fig:RS}, again doing the same time average as in LRC.

\begin{equation}
    RS=<\tilde{v_r}\tilde{v_\theta}>
    \label{eqn:RS}
\end{equation}

\subsubsection{Radial electric field ($E_r$)\newline}
\begin{figure}[H]
    \centering
    \textbf{\hspace*{-5mm}}\includegraphics[width=7.7cm]{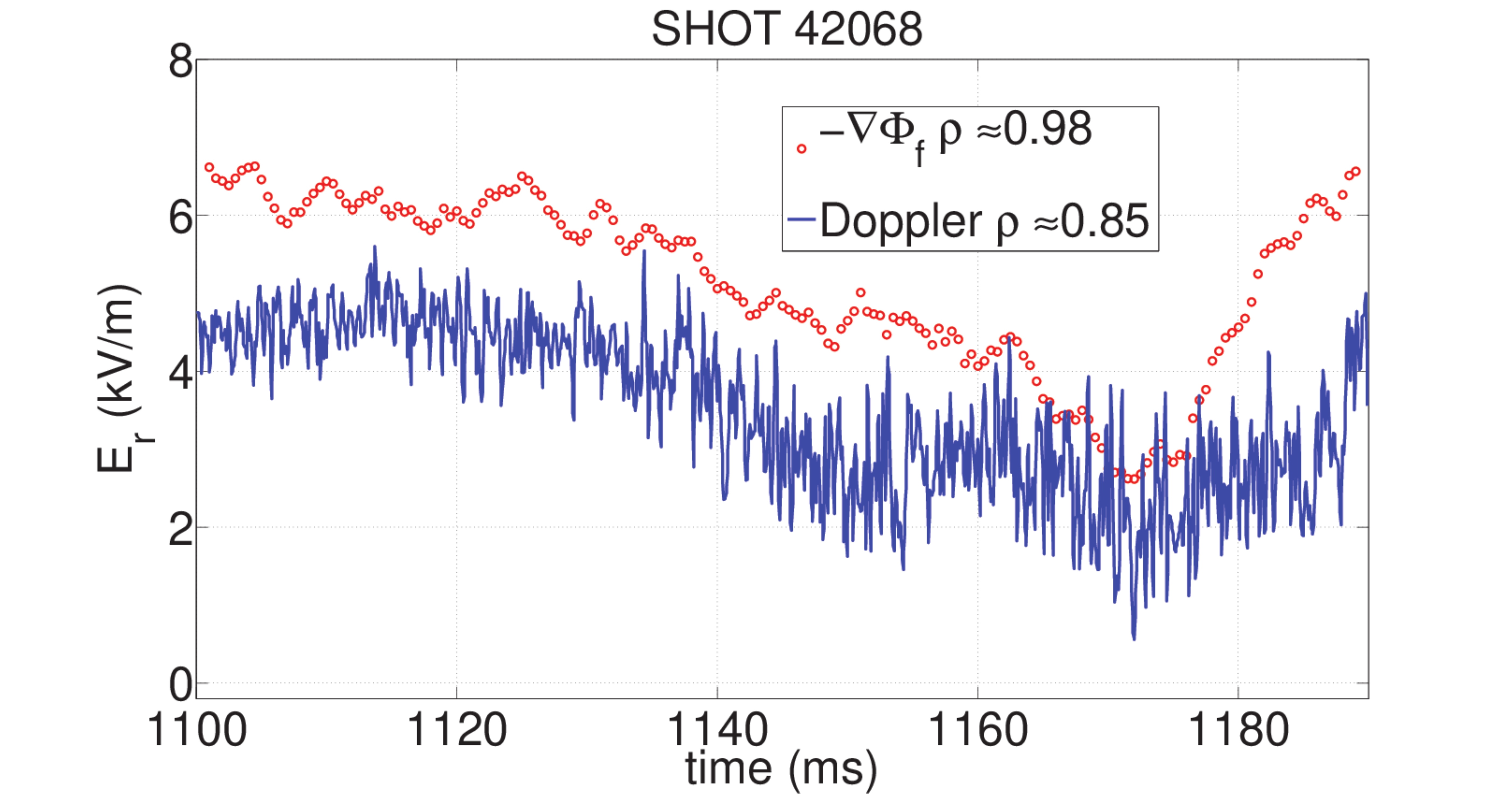}
    \caption{Measurements of radial electric field by Doppler reflectometry and by gradients of floating potential when approaching to the electron-ion transition with a density ramp (see figure \ref{fig:trans_example}).}
    
    \label{fig:doppler}
\end{figure}

Comparisons between gradient in floating potential (measured in the point where $\nabla RS$ is calculated, $\rho\approx0.98$) with radial electric fields measured by Doppler reflectometry (measured at $\rho\approx0.85$) have been done \cite{Happel}. These two measurements show qualitative agreement (i.e. changes in the range of few kV/m with biasing and a reduction in the $E_r$ value when approaching to the electron-ion root transition as can be seen in Fig. \ref{fig:doppler}). Therefore, in the present experimental conditions, $-\nabla\Phi_f$ can be used as a proxy of the radial electric field.\\
\\
For the calculation of radial electric field from floating potential, we have used the low frequencies of the $\phi_f$ signals, from 0 to 2 kHz. Making the gradient in radial direction of this signals in probe D, 4 measurements of radial electric field are obtained and averaged to get a measurement in the geometric point where the $\nabla RS$ and LRC are calculated.\\

\subsection{Simulation code}
For the study of the influence on ion orbit losses of radial electric fields in the outer part of the plasma, a simulation have been run with ISDEP \cite{ISDEP}. ISDEP is a Monte Carlo code that launches independent ions in a background plasma and follows their trajectories (which are governed by the magnetic and electric fields and random collisions with bulk electrons and ions) until it escapes beyond the last closed magnetic flux surface.\\
\\
Simulations were carried out using density, temperature, potential and electric field profiles that qualitatively resemble experimental profiles in electron and ion root plasma scenarios, and with the potential and the electric field profiles modified for $\rho>0.8$, with constant electric field values from zero to $\pm$8 kV/m (with the sign corresponding to the electron or ion root phase). Around $3\times 10^{4}$ particles were launched with original position between $\rho=0.85$ and $\rho=0.95$, with random pitch evenly distributed and random Maxwellian energy distribution centered in 100 eV.\\
\\
This code calculates the percentage of particles that remains confined at several certain times (t=0.1 ms, 0.2 ms, etc.), like in figures \ref{fig:Pers_ECRH} and \ref{fig:Pers_NBI}. The dependence of this persistence on the radial electric field in the outer part of the plasma has been studied.

\section{Results}
\begin{figure}[H]
    \centering
    \textbf{\hspace*{-5mm}}\includegraphics[width=8cm]{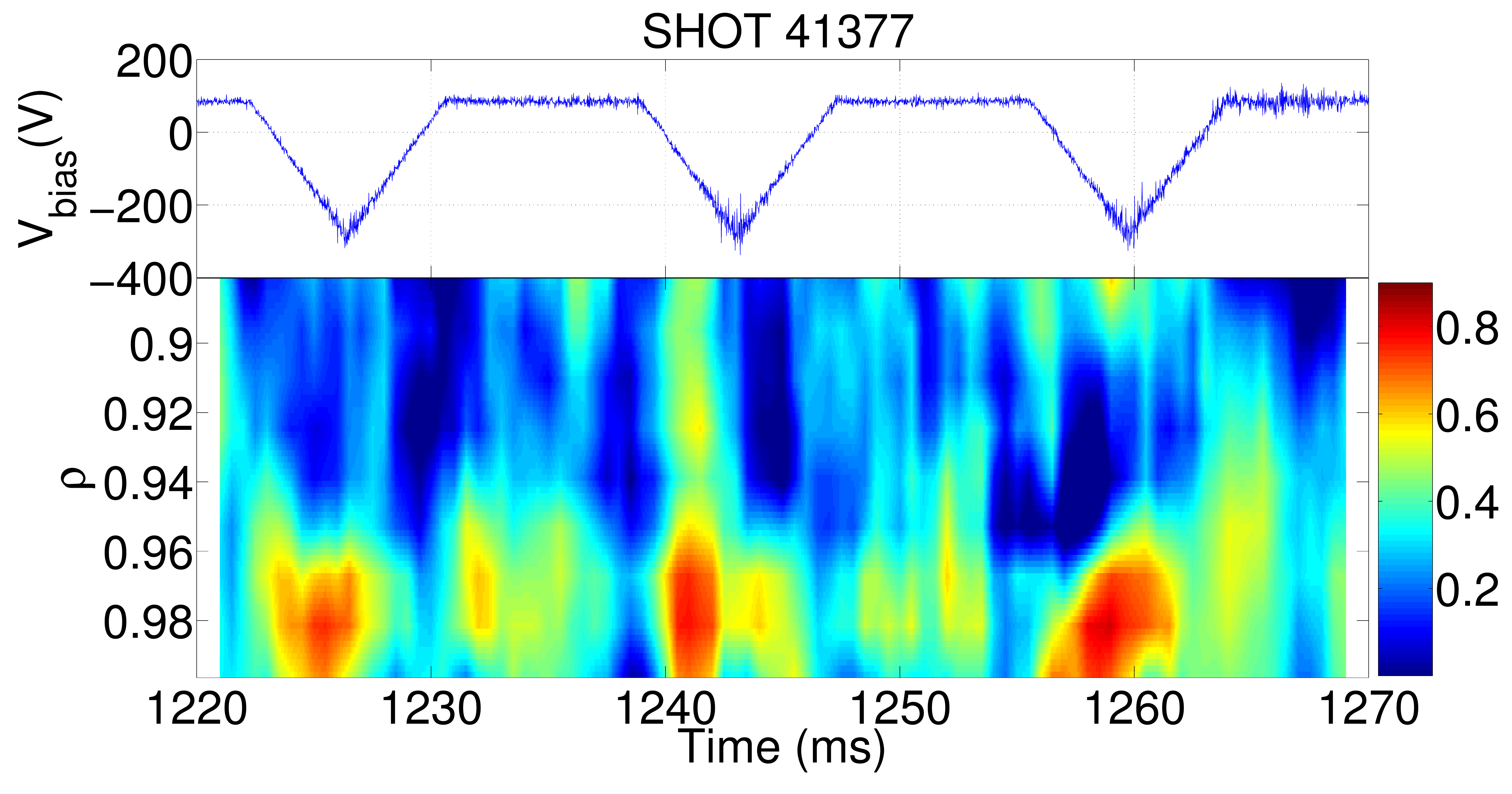}
    \caption{Effect of biasing (upper plot) in LRC measurement in an NBI shot (down plot).}
    \label{fig:intro}
\end{figure}

Figure \ref{fig:intro} shows the influence of edge biasing in the amplitude of LRC measured in NBI plasma scenarios. In agreement with previous results \cite{Pedrosa,Wilcox, Manz, Xu} the amplitude of LRC is modulated by edge $E_r$.\\
\\
It is important to stress that the TJ-II unique experimental set-up allows to measure simultaneously and at the same radial location (with resolution $\Delta\rho=0.02$), the amplitude of LRC, the gradient of Reynolds stress and $E_r$ values in ion root (section 3.1), electron root (section 3.2) and in the proximity of the electron-ion root transition (section 3.3) scenarios.
\\

\subsection{Ion root phase results}
\begin{figure}[]
    \centering
    \begin{subfigure}
    \centering
    \textbf{\hspace*{-1cm}}\includegraphics[width=8.5cm]{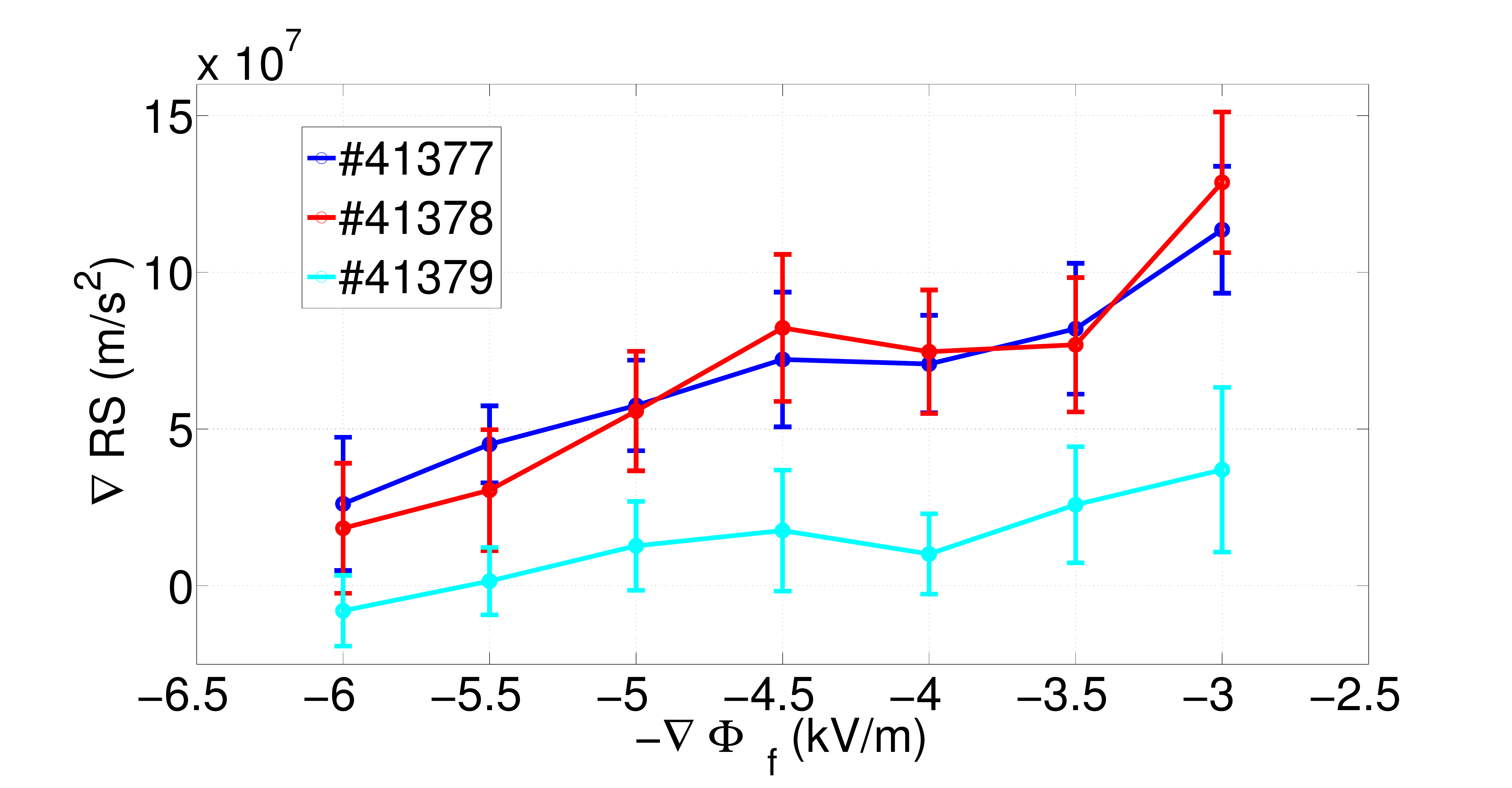}
   \end{subfigure}
    \begin{subfigure}
    \centering
     \textbf{\hspace*{-1cm}}\includegraphics[width=8.5cm]{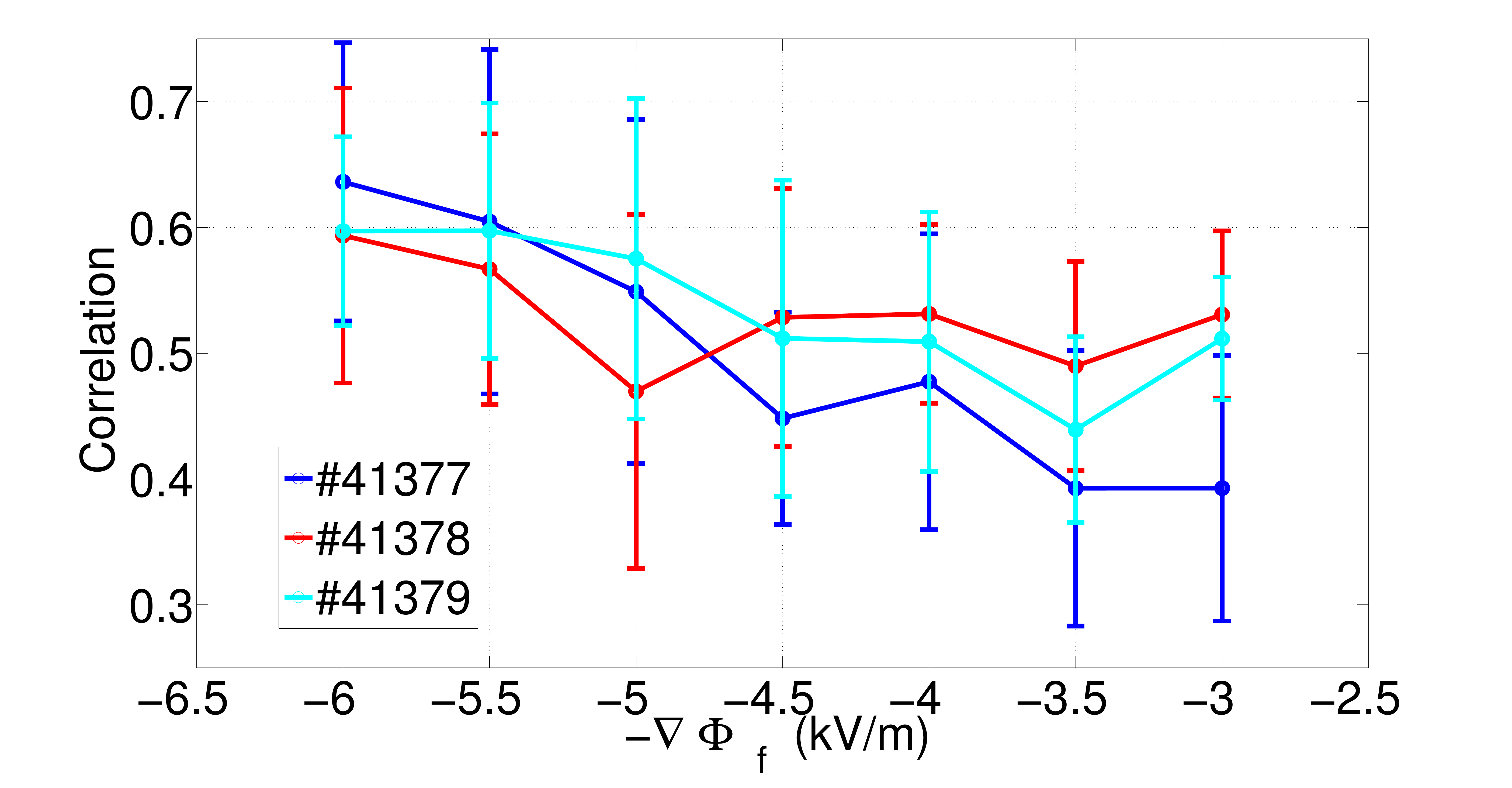}
   \end{subfigure}
 \caption{Ion root phase: Top: $\nabla RS$ vs $-\nabla\Phi_{f}$ at $\rho\approx0.98$. Bottom: LRC vs$-\nabla\Phi_{f}$ at $\rho\approx0.98$.}
    \label{fig:NBI_results}
   
\end{figure}

\begin{figure}
    \centering
    \begin{subfigure}
    \centering
    \textbf{\hspace*{-0.5cm}}\includegraphics[width=8.5cm]{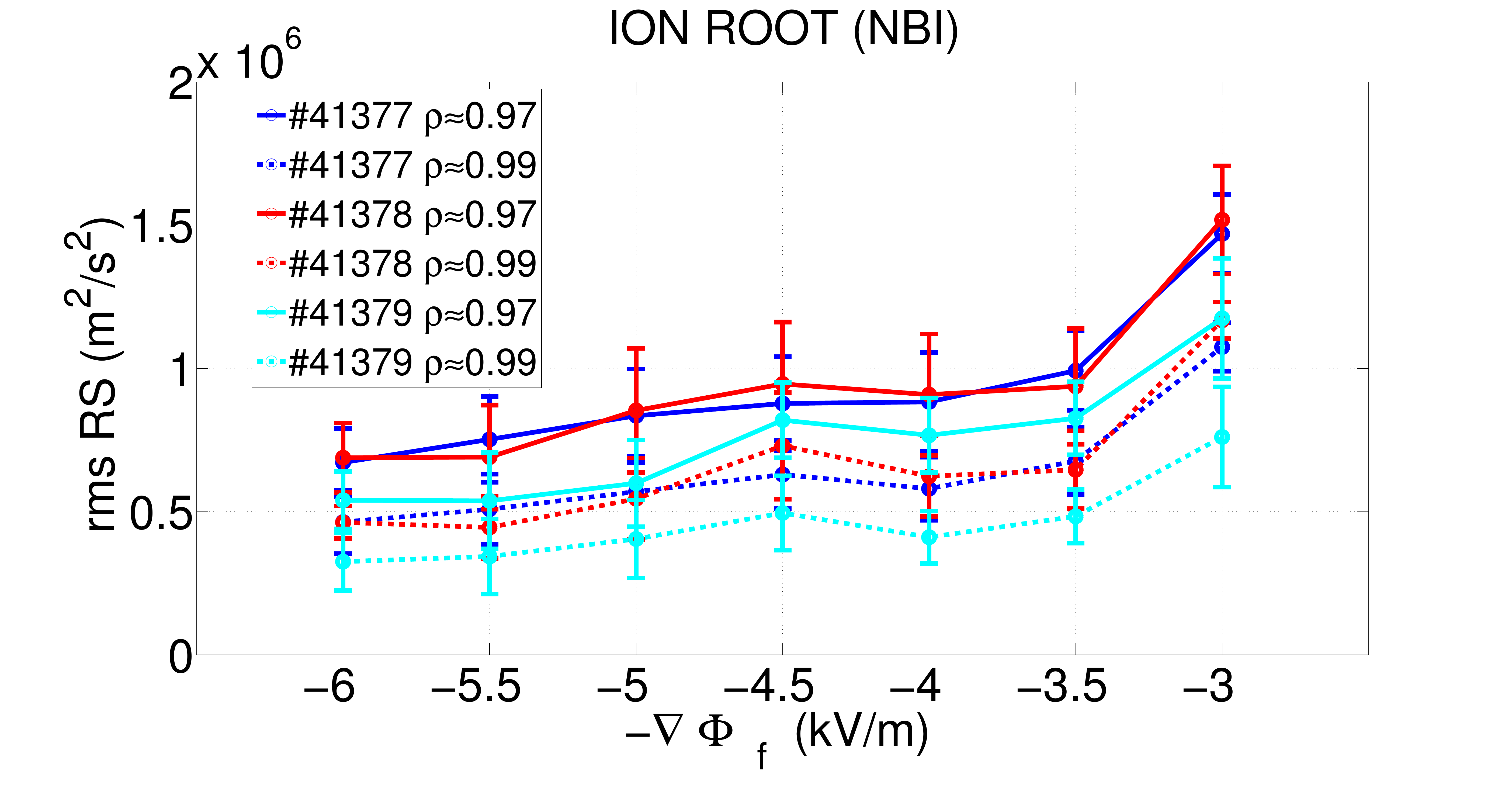}
   \end{subfigure}
    \begin{subfigure}
    \centering
     \textbf{\hspace*{-0.5cm}}\includegraphics[width=8.5cm]{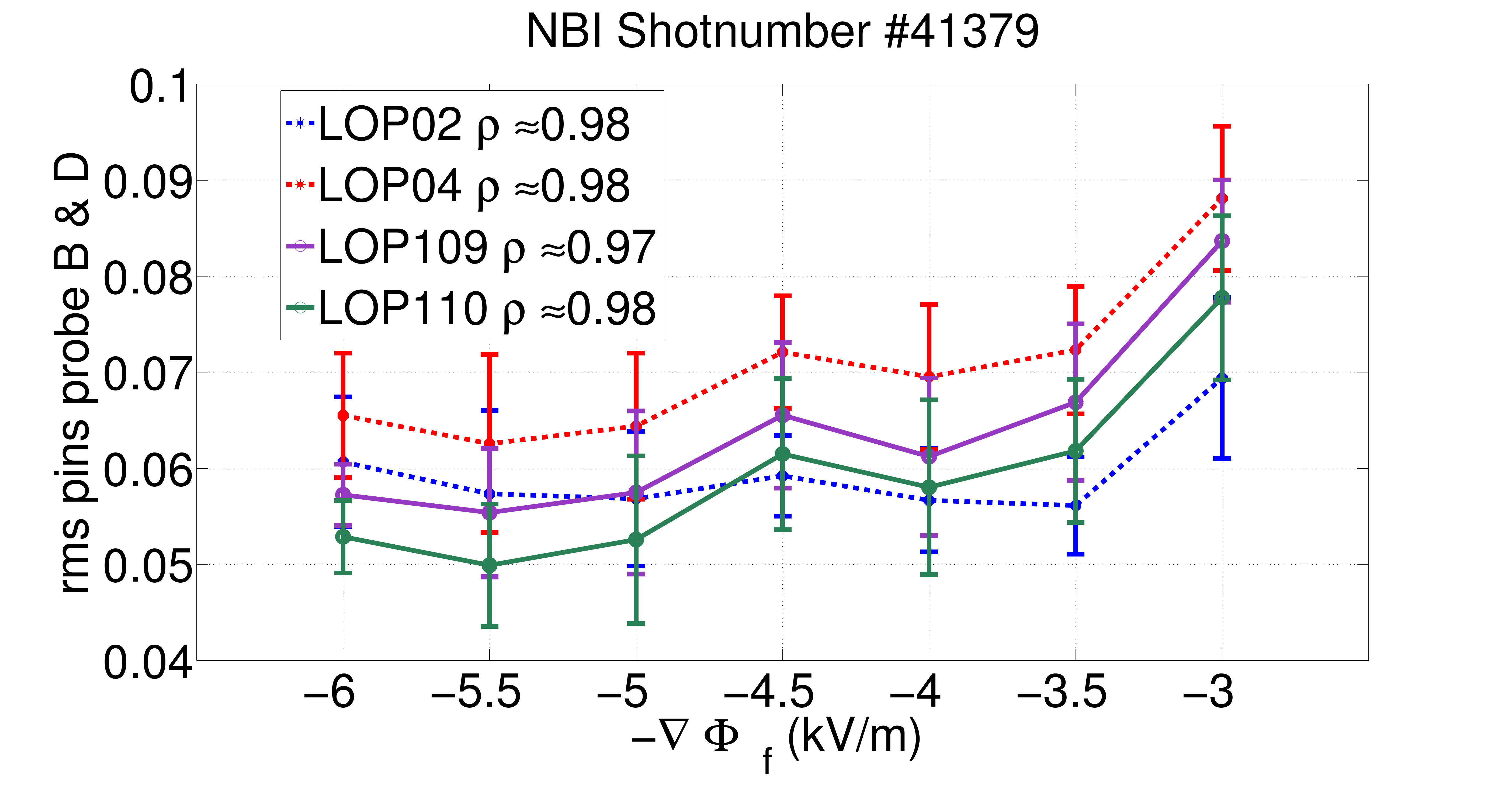}
   \end{subfigure}
 \caption{Ion root phase: Top: rms of RS at different radial positions vs $-\nabla\Phi_{f}$. Bottom: rms of $\phi_f$ vs $-\nabla\Phi_{f}$ for probe D (LOP02 and LOP04) and probe B (LOP109 and LOP110).}
    \label{fig:rms_figures}
   
\end{figure}

The results on the NBI plasmas were obtained with a constant average density of around $1.75\times10^{19}m^{-3}$ and core electron and ion temperatures in the range of 300 eV and 130 eV respectively.\\
\\
The top figure \ref{fig:NBI_results} shows the reduction of $\nabla RS$ when the magnitude of the gradient in floating potential (as a proxy of $E_r$) increases. The observed reduction in the gradient of Reynolds stress as $E_r$ increases is experimentally correlated with a slight reduction in the level of fluctuations of Reynolds stress measurements (rms of RS) and in the plasma floating potential fluctuations (rms of $\phi_f$), as can be observed in Fig. \ref{fig:rms_figures}. The reduction of plasma fluctuations is observed both in probe 1 (where gradients in Reynolds stress are not available) and 2, which points out that the observed tendency describes a global phenomenon.\\
\\
Meanwhile, bottom figure \ref{fig:NBI_results} presents an slight increment in LRC amplitude (despite the magnitude of error bars produced by the dispersion of our average measurements for several cycles of biasing perturbation, as it is shown in figure \ref{fig:intro}) with the increasing magnitude of the electric field, which implies that ZF activity increases with the radial electric field magnitude. This result agrees with previous investigations done in the device \cite{Pedrosa}.

\begin{figure}[]
    \centering
    \begin{subfigure}
    \centering
   \textbf{\hspace*{-0.4cm}}\includegraphics[width=8.5cm]{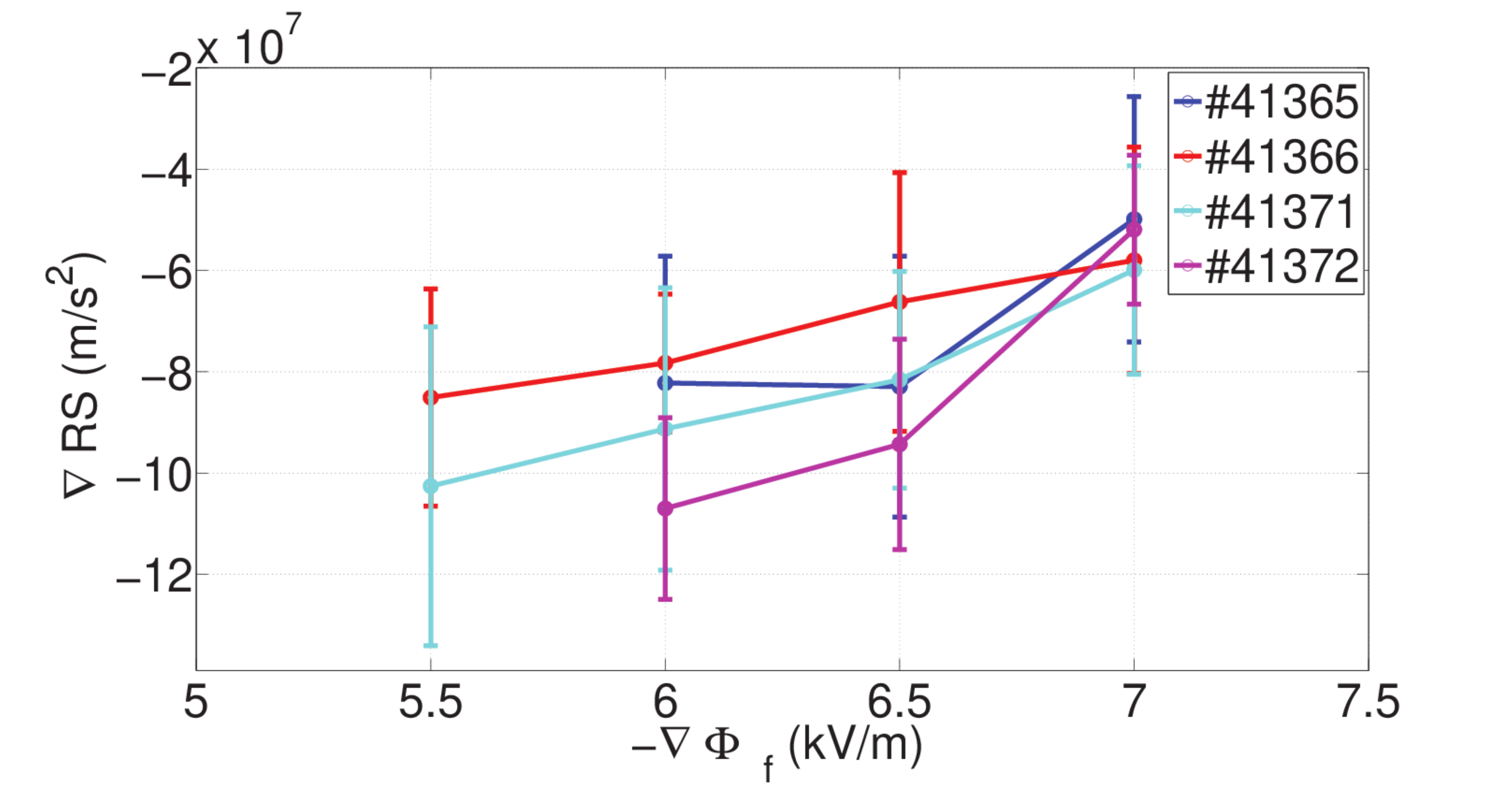}
   \end{subfigure}
    \begin{subfigure}
    \centering
    \textbf{\hspace*{-0.4cm}}\includegraphics[width=8.5cm]{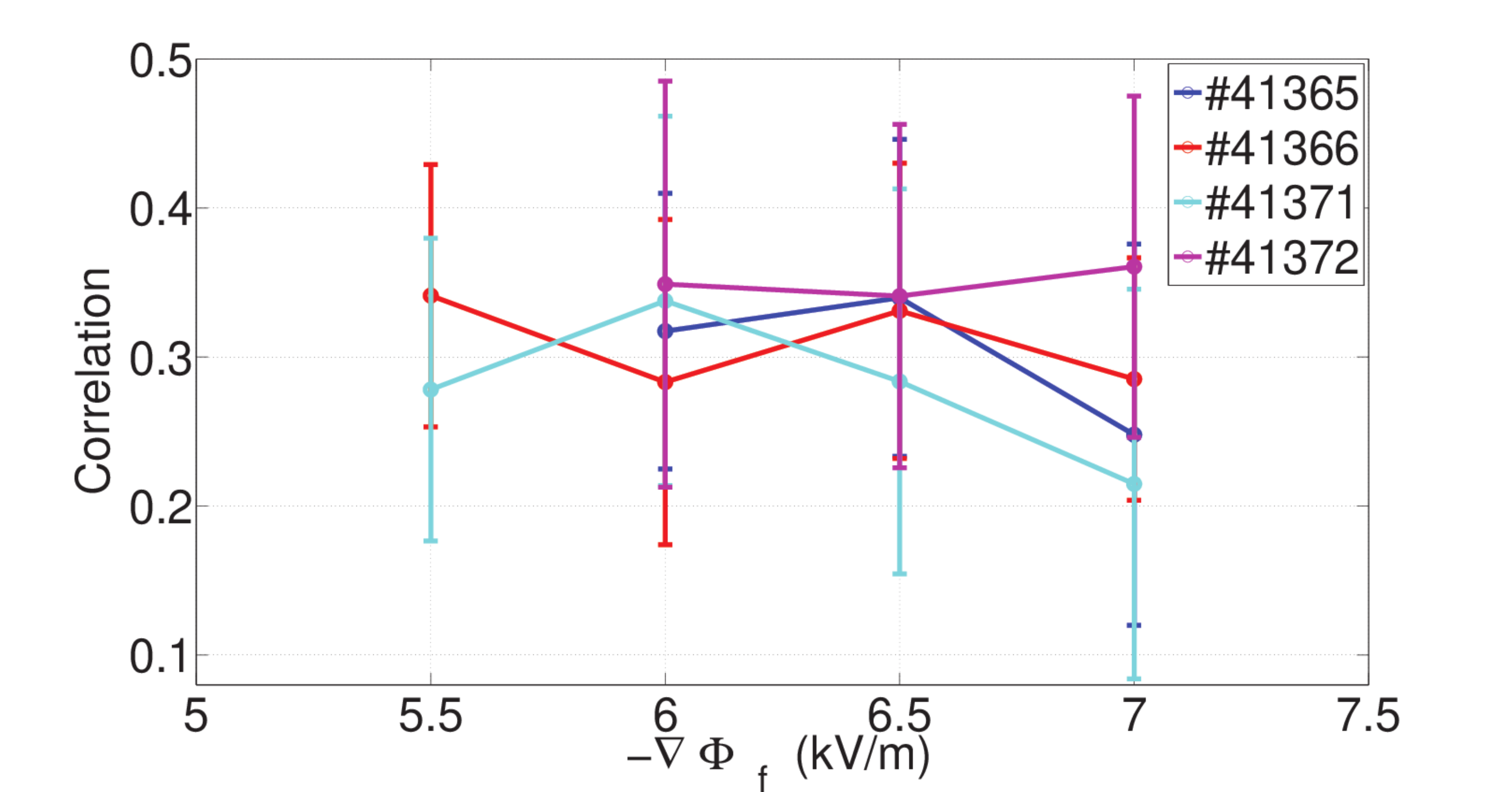}
   \end{subfigure}
 \caption{Electron root phase: Top: $\nabla RS$ vs $-\nabla\Phi_{f}$ at $\rho\approx0.98$. Bottom: LRC vs$-\nabla\Phi_{f}$ at $\rho\approx0.98$. }
    \label{fig:ECRH_results}
   
\end{figure}

\subsection{Electron root phase results}
The results on the ECRH plasmas, with a constant density of around $0.4\times10^{19}m^{-3}$, core electron and ion temperatures in the range of  1 keV and 70 eV respectively, are shown in figure \ref{fig:ECRH_results}.\\
\\
It can be seen the slight reduction of the absolute value of $\nabla RS$ when the magnitude of the electric field increases, as in the ion root scenario. It should be noted that the sign of $\nabla RS$ reverses from positive to negative when the sign of $E_r$ changes from electron to ion root (see figures \ref{fig:NBI_results} and \ref{fig:ECRH_results}). 
\\
\\
There is also, as in ion root, a reduction in the fluctuation level in $\phi_f$ measured by probe 1 and 2, with a decrease in the absolute value of RS. Meanwhile, the amplitude of LRC is not affected by $E_r$ changes within experimental uncertainties. 

\subsection{Electron-ion transition} 
\begin{figure}[H]
    \centering
    \textbf{\hspace*{-1.1cm}}\includegraphics[width=8.5cm]{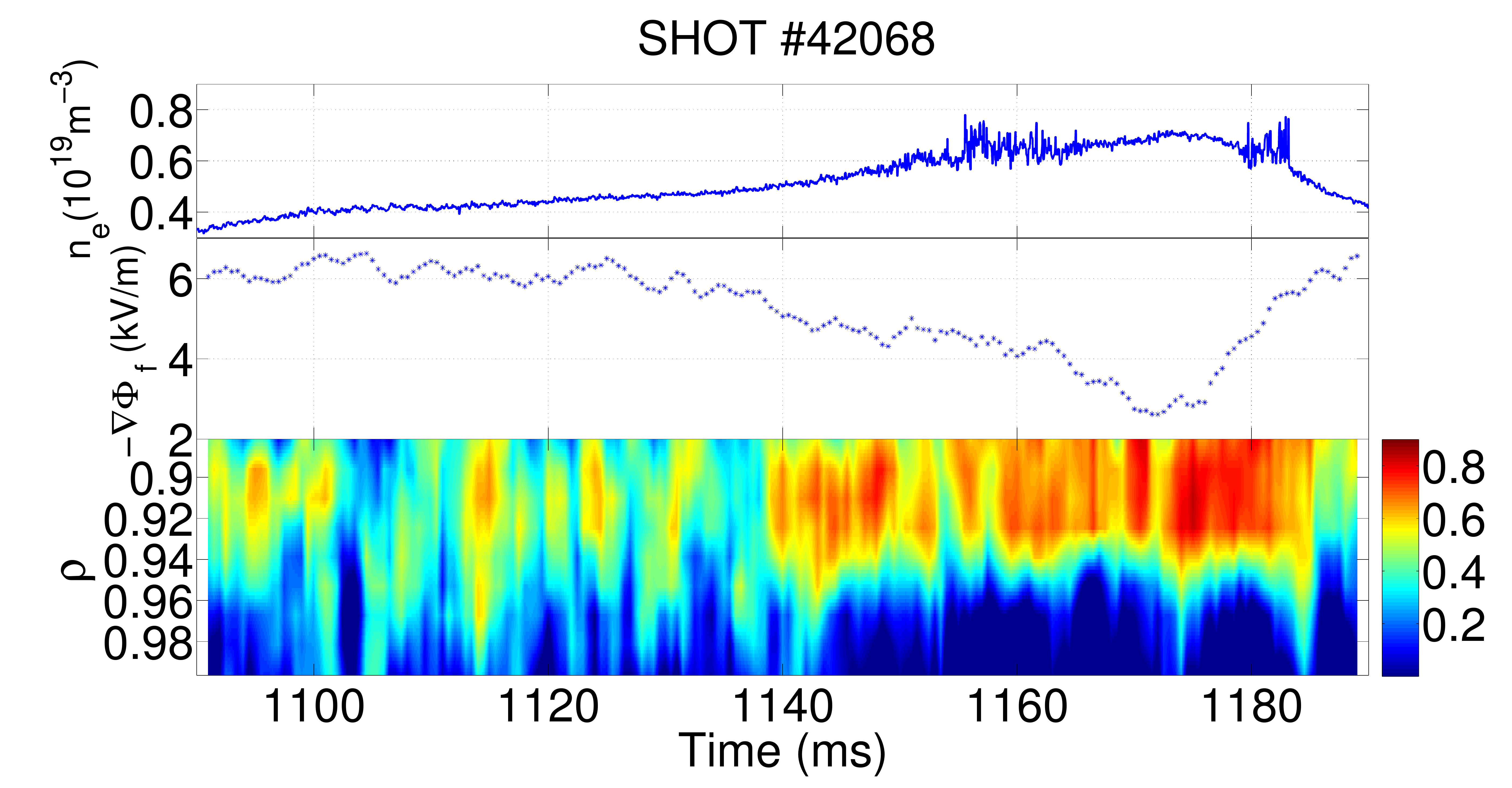}
    \caption{Up: $n_e$ ramp up to around the transition density. Middle: $-\nabla \phi_f$ in $\rho\approx0.92$. Bottom: LRC during the transition.}
    \label{fig:trans_example}
\end{figure}

\begin{figure}
    \centering
    \begin{subfigure}
    \centering
   \textbf{\hspace*{-0.5cm}}\includegraphics[width=8.5cm]{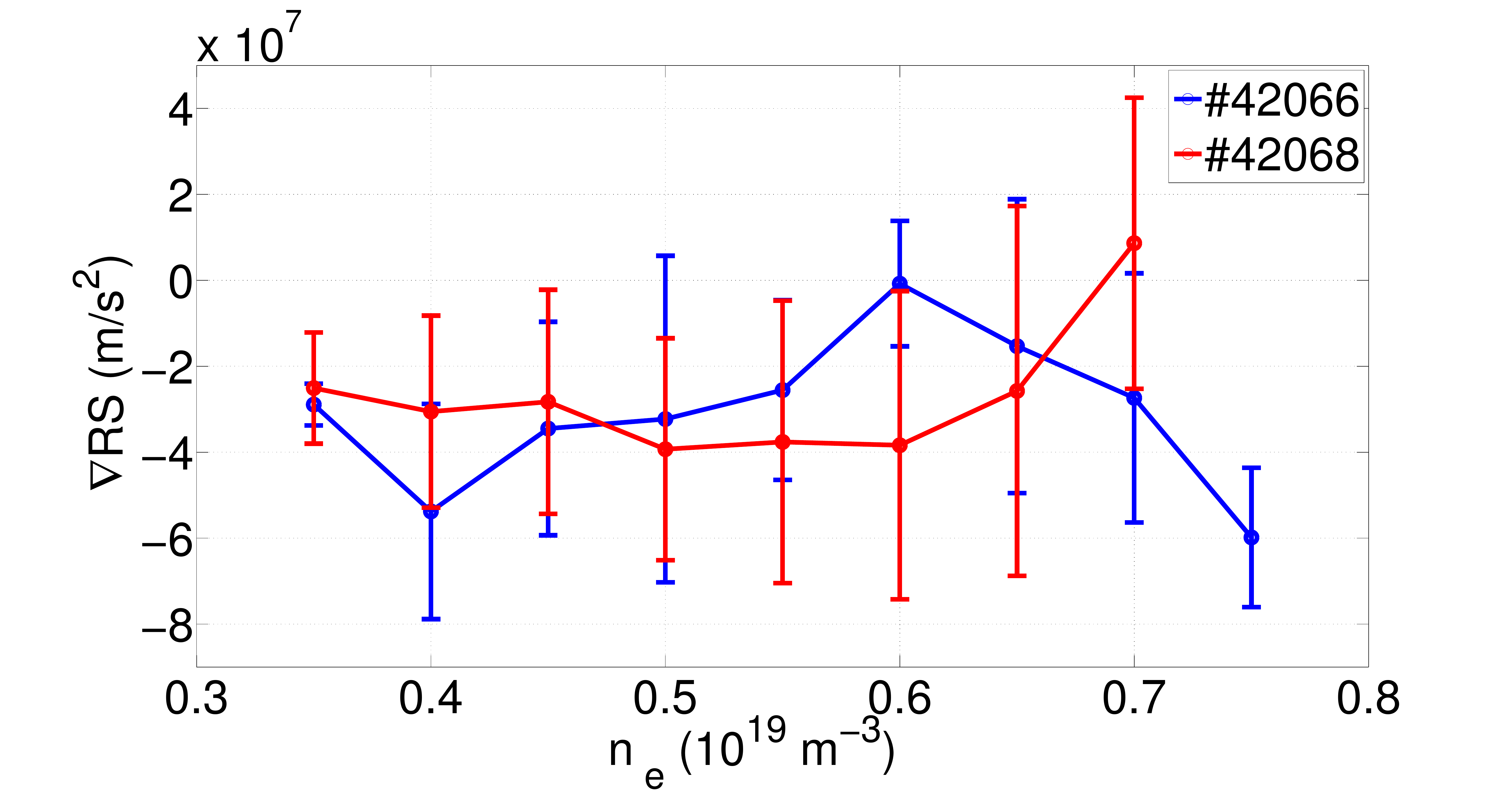}
   \end{subfigure}
   
    \begin{subfigure}
    \centering
    \textbf{\hspace*{-0.5cm}}\includegraphics[width=8.5cm]{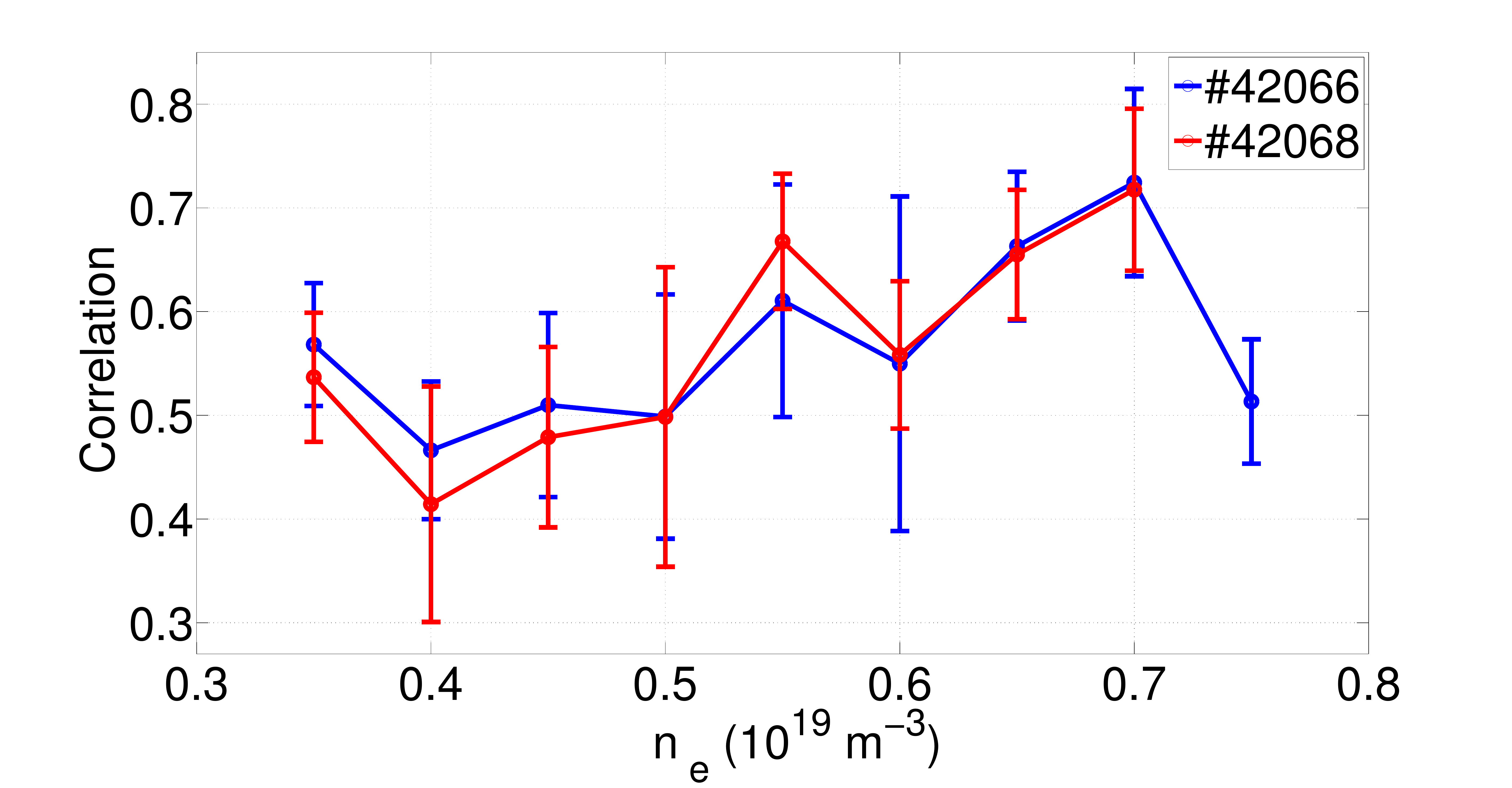}
   \end{subfigure}
 \caption{Electron-Ion transition: Top: $\nabla RS$ vs $n_e$ at $\rho\approx0.92$. Bottom: LRC vs$n_e$ at $\rho\approx0.92$. }
    \label{fig:transition_results}
   
\end{figure}

In this section, the dependency of $\nabla RS$ and LRC is presented as plasma density approaches the transition between electron and ion root (as can be seen in figure \ref{fig:trans_example}). In this transition, the gradient of floating potential changes from positive values (electron root) towards negative ones (ion root) (Fig. \ref{fig:doppler} and \ref{fig:trans_example}) and an increment of the LRC happens when plasma gets close to the transition density ($n_e\approx0.6\times10^{19}m^{-3}$).
Bottom figure \ref{fig:transition_results} shows how the LRC clearly increases when we get closer to the transition density, whereas, for the $\nabla RS$, does not show any correlation with the increment of density (related with a reduction in $E_r$) within experimental uncertainties.

\subsection{Influence of $E_r$ in particle losses} 
\begin{figure}[]
    \centering
    \textbf{\hspace*{-0.5cm}}\includegraphics[width=8.5cm]{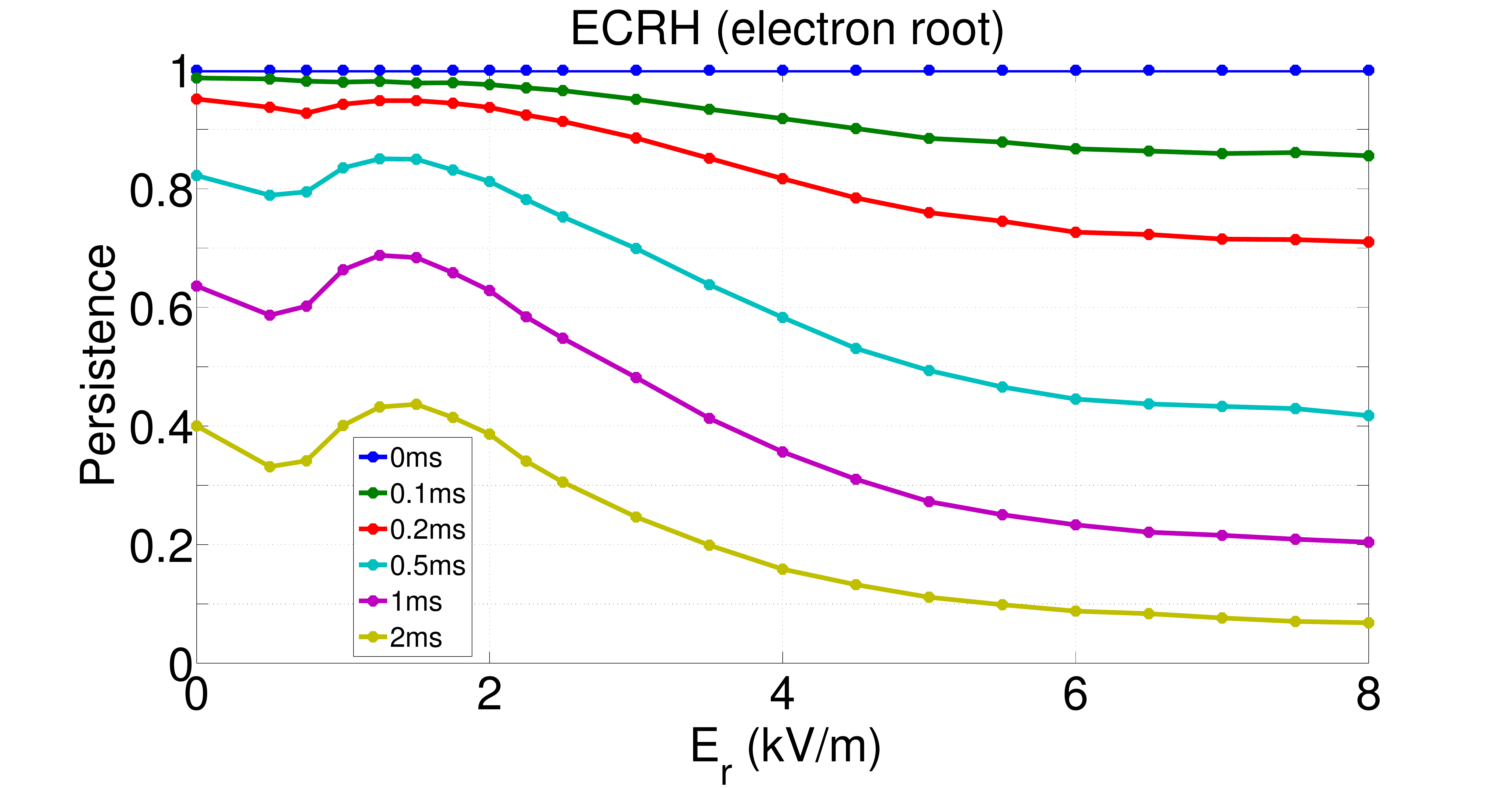}
    \caption{Persistence obtained from ISDEP code for the ECRH phase}
    \label{fig:Pers_ECRH}
\end{figure}

\begin{figure}[]
    \centering
    \textbf{\hspace*{-0.5cm}}\includegraphics[width=8.5cm]{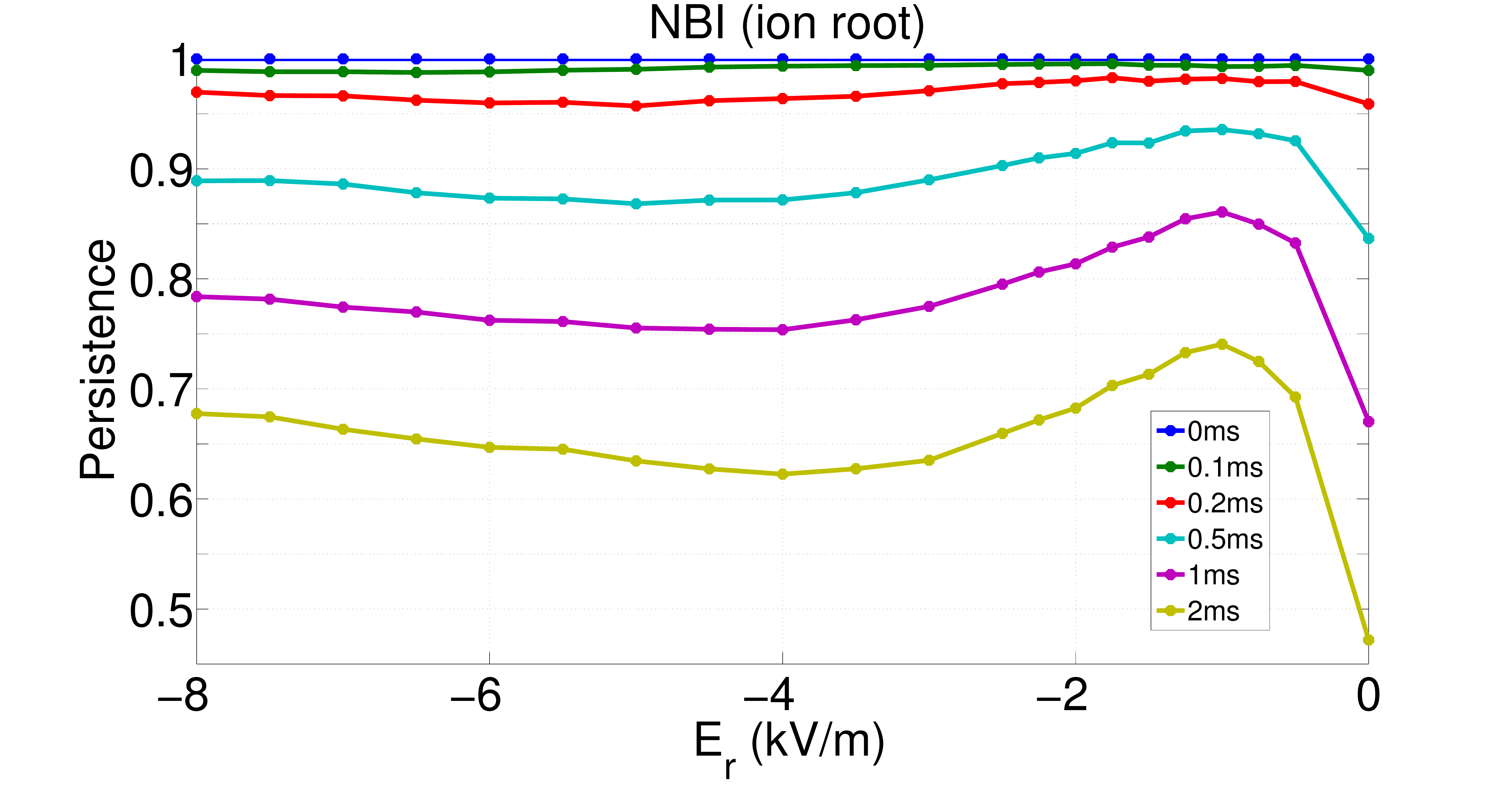}
    \caption{Persistence obtained from ISDEP code for the NBI phase}
    \label{fig:Pers_NBI}
\end{figure}

Figures \ref{fig:Pers_ECRH} and \ref{fig:Pers_NBI} show the level of particle persistence calculated by means of the ISDEP code for ECRH and NBI plasma scenarios.\\
\\
It can be seen that the persistence shows qualitative changes for variations of around 1 kV/m in electric field in both scenarios.\\
\\
In both plasma regimes particle persistence decreases as time scale increases (particularly beyond 0.2 ms). In plasmas with negative $E_r$ (i.e. NBI-like) particle persistence strongly increases for $E_r$ values in the range 0-1 kV/m, decreasing for $E_r>1\>kV/m$ and reaching a minimum for $E_r\approx4\>kV/m$. In plasmas with positive $E_r$ (i.e. ECRH-like) particle persistence decreases for $E_r$ greater than 1.5 kV/m, having some fluctuations for low electric fields. These dependencies for low electric fields magnitude depend on particle energy, pointing out the competition between $E\times B$ and $B\times\nabla B$ drift velocities. It is important to notice that for electric fields $E_r=0\>kV/m$, the persistence is really similar in NBI and ECRH simulation.
\\ 

\section {Discussion and conclusions}
The results presented here were obtained thanks to a unique experimental set-up that allows the simultaneous measurement of $\nabla RS$, LRC amplitude and radial electric field.\\
\\
The order of magnitude of the $\nabla RS$ ($10^7 m/s^2$) is high enough to produce turbulent forces comparable to the neoclassical damping forces of poloidal flows (assuming values of plasma viscosity $\mu\approx10^4-10^5 s^{-1}$ \cite{Velasco}).\\
\\
However, its dependency with $E_r$ and its correlation with the amplitude of LRC is totally different for the three different plasma scenarios investigated: Decreasing $\nabla RS$ and increasing LRC amplitude with the increment of $E_r$ magnitude in ion root phase, also decreasing $\nabla RS$ but without any dependence of LRC amplitude within experimental uncertainty with increasing magnitude of $E_r$ in electron root phase and an increment of LRC amplitude but not a clear dependence of $\nabla RS$ withing experimental uncertainty with the increment of $n_e$ up to transition density (which is related with a change of $E_r$) in the electron to ion root transition.\\
\\
It is also showed with ISDEP simulations that changes of radial electric field of around 1 kV/m in the edge of the plasma, as the ones produced in the experimental shots, have an impact in neoclassical orbit losses (in TJ-II).\\
\\
This results indicate that an interplay between turbulent and neoclassical mechanisms should be taken into account for a complete understanding of the effects of radial electric fields in the mechanisms of production and suppression of edge Zonal Flows \cite{Velasco, Alexey}.

\section*{ACKNOWLEDGEMENTS}

This research has been supported by the grant '\textit{Fomento a la investigacion}' of Autonomous University of Madrid (UAM).

\end{document}